\documentclass[12pt]{article}
\usepackage{amsmath} 
\usepackage{epsfig} 
\input{epsf} 
\newcommand\as{\alpha_{{S}}} 
\newcommand\f[2]{\frac{#1}{#2}} 
\def\la{\lambda} 
\def\b0{b_0}

\def\beq{\begin{equation}} 
\def\eeq{\end{equation}} 
\def\beeq{\begin{eqnarray}} 
\def\eeeq{\end{eqnarray}} 
 
\def\to{\rightarrow}
 
\def\nn{\nonumber}

\def\GE{\gamma_E}

\def\astpi{{\alpha_S \over 2 \pi}}

\def\pt{p_{T}}
\def\ptsq{p_{T}^2}
\def\mT{m_{T}}

\def\sig{\sigma}
\def\sigh{\hat{\sigma}}
\def\sh{\hat s}

\def\yH{y_{H}}
\def\yT{y_{T}}
\def\yTh{\hat y_{T}}

\def\gev2{{\rm GeV}^2}
\def\nn{\nonumber}

\def\rar{\rightarrow}

\def\lapproxeq{{\ \lower 0.6ex \hbox{$\buildrel<\over\sim$}\ }}
\def\gapproxeq{{\ \lower 0.6ex \hbox{$\buildrel>\over\sim$}\ }}

\def \be  {\begin{equation}}
\def \ee  {\end{equation}}
\def \ba  {\begin{eqnarray}}
\def \ea  {\end{eqnarray}}
\def \baa {\begin{eqnarray*}}
\def \eaa {\end{eqnarray*}}
\def \bb  {\begin{thebibliography}}
\def \eb  {\end{thebibliography}}

\begin{document} 

\begin{titlepage}
\renewcommand{\thefootnote}{\fnsymbol{footnote}}
\begin{flushright}
\vspace*{-3cm}
BNL-NT-05/43\\
RBRC-573\\
TTP05-23\\
SFB/CPP-05-75\\
hep-ph/0511205
\end{flushright}

\par 
\vspace{5mm}

\begin{center}
{\Large \bf
Threshold resummation\\[.2em]
for high-transverse-momentum Higgs production\\[.5em]
at the LHC}
\end{center}
\par \vspace{2mm}
\begin{center}
{\sc Daniel de Florian${}^{(a)}$, Anna Kulesza${}^{(b)}$
and
Werner Vogelsang${}^{(c)}$}\\

\vspace*{8mm}

{\it \small
${}^{(a)}$Departamento de F\'\i sica, FCEYN, Universidad de Buenos Aires,\\
(1428) Pabell\'on 1 Ciudad Universitaria, Capital Federal, Argentina\\
\vspace{2mm}

${}^{(b)}$Institut f\"ur Theoretische Teilchenphysik, Universit\"at Karlsruhe 
\\ D-76128, Karlsruhe, Germany\\
\vspace{2mm}

${}^{(c)}\,$Physics Department and RIKEN-BNL Research Center, \\
Brookhaven National Laboratory, Upton, NY 11973, U.S.A.\\
}
\end{center}

\vspace*{2.5cm}

\begin{center} {\bf Abstract} \end{center}
\begin{quote}
We study the resummation of large logarithmic QCD corrections for the
process $pp \rar H+ X$ when the Higgs boson $H$ is produced at high
transverse momentum. The corrections arise near the 
threshold for partonic reaction and originate from soft gluon emission.  
We perform the all-order resummation at next-to-leading logarithmic
accuracy and match the resummed result with the next-to-leading order
perturbative predictions. The effect of resummation on the Higgs transverse
momentum distribution at the LHC is discussed.
\end{quote}
\par

\vspace*{1.5cm}
\begin{flushleft}
November 2005
\end{flushleft}
\end{titlepage}

\def\thefootnote{\arabic{footnote}}
\setcounter{footnote}{0}

\section{Introduction}
\label{sec:intro}

To date, the Higgs boson, 
responsible for the mechanism of the electroweak symmetry breaking, 
remains the only undetected ingredient of the Standard Model. The
search for the Higgs boson is one of the highest
priorities for both the CERN Large Hadron Collider (LHC) and Fermilab 
Tevatron physics programs \cite{Gunion:1989we}.
Direct searches at LEP and fits to electroweak precision data
 indicate that it might be light, with a lower bound of  $114.4$~GeV
 \cite{Barate:2003sz} and an upper limit of 
$m_H < 260$~GeV at $95\%$ CL \cite{Group:2004qh}.

The detection of the Higgs boson in the low mass range, $m_H <
140$~GeV, though feasible, will not be 
a simple task even at the LHC \cite{atlascms,lhcupdate}. 
The dominant production channel at hadron colliders is the gluon 
fusion, mediated at lowest order in the SM by a heavy (mainly top) 
quark loop. In the considered mass range, experimental searches at 
the LHC will concentrate on the rare two-photon decay mode $H \rar 
\gamma + \gamma$. In the absence of any constraints imposed on the events, 
the bulk of the cross section will be at relatively low transverse
momenta of the photon pair, where the background is large. 
Thus, despite the high production cross section, the detection of 
the signal is considered a challenging task.  

A possible way to improve the signal 
significance for Higgs discovery in the considered mass range is
to study the less inclusive $\gamma+\gamma{\rm \ + \ jet(s)}$ 
final states\footnote{The 
study of Higgs production in association with a jet
was first suggested in the context 
of improving the $\tau$ reconstruction 
in the $\tau^+\tau^-$ decay channel \cite{highptLOellis}.}, which offer
several advantages \cite{abdullin}. In this case the photons are more 
energetic than for the inclusive channel, and 
the reconstruction of the jet in the calorimeter allows a
more precise determination of the interaction vertex, improving the 
efficiency and mass resolution. Furthermore, the existence of a jet 
in the final state allows for a new type of event selection and a 
more efficient background suppression. Also theoretical
considerations make the process appear more favorable regarding its
background: while for the fully inclusive channel the $gg\to \gamma\gamma$
background contribution that first enters at NNLO is as sizable as the Born 
cross section for the Higgs production \cite{back} and thus complicates
the organization of the perturbative calculation, it is significantly 
suppressed and less relevant at large transverse momentum of the photon pair 
\cite{ggbackground}. One therefore expects that the background is under better 
theoretical control than that for the inclusive cross section.

Quite generally, on the theoretical side, signal and background cross 
sections need to be calculated with the highest possible accuracy, 
minimizing the theoretical uncertainties. In the case of Higgs boson 
production cross-sections, the main theoretical uncertainties
come from two sources: the parton distributions, and the use of
QCD perturbation theory for the partonic hard-scattering. 
Typical uncertainties in the relevant parton distributions are only 
of the order of a few percent, and further improvements are expected from
new data that will become available from the usual Standard Model processes 
at the LHC. Regarding the status of perturbation theory for the partonic 
cross-sections for Higgs production, a very slow convergence of 
the perturbative 
expansion was observed for the fully inclusive Higgs production
cross section, for which the next-to-leading-order (NLO) contributions were
found to be as large as the leading-order (LO) 
term \cite{Dawson:1991zj,Spira:1995rr}. 
Consequently, an enormous effort was devoted to obtaining the next
order (NNLO) in perturbative QCD, which turned out to be under 
better control, albeit still sizeable \cite{NNLOtotal}. 
Due to the high complexity of the
 calculations (the lowest order is already a one-loop process because
a top quark loop is required to couple the gluons to the Higgs), the
 results for the NNLO corrections were obtained
 in the large-top-mass $m_t$ limit, i.e. $m_t \rar \infty$. In this
 limit the top quark loops may be replaced by point-like vertices, and
 the Feynman rules are given by an effective Lagrangian. At NLO, the
 method is known to provide a very good approximation of the exact
 result for $m_H < 2 m_t$~\cite{Kramer:1996iq}.  

The origin of the large size of the higher order contributions to the 
perturbative partonic cross sections can be traced to the presence of 
large logarithmic terms, referred to as ``threshold logarithms''.
These result from the emission of soft and collinear gluons near the edges 
of phase space. It turns out that the threshold logarithms, along with terms 
from purely virtual corrections, account for more than 90\% of the total 
cross section for inclusive Higgs production  at the 
LHC~\cite{Catani:2001ic,Harlander:2001is}. The most important
(leading) logarithms at the $n$th order in perturbation theory are of the form
$\as^k \left(\ln^{2k-1}(1-z)/(1-z)\right)_+$, where $1-z=m_H^2/\sh$, 
$\sh$ being the partonic center-of-mass energy. Sufficiently close to the 
partonic threshold at $\sh=m_H^2$ or $z=1$, where the initial state
partons have just enough energy to produce the Higgs boson, fixed-order
approximations of the partonic cross sections are bound to fail, no matter
how small the coupling constant. These effects of multiple soft gluon emission 
can, however, be taken into account to all orders in perturbation 
theory by performing a resummation of the threshold logarithms. For 
inclusive Higgs boson production, the resummation is completely 
known to the next-to-next-to-leading logarithmic (NNLL)
accuracy \cite{Catani:2003zt} \footnote{Many of the ingredients needed to 
perform the resummation to complete N$^3$LL accuracy became available recently 
\cite{Moch:2005ky,Laenen:2005uz,Idilbi:2005ni}.}. 
Thanks to the interplay of the partonic 
cross sections with the parton distributions, threshold resummation 
considerably improves the predictive power of the theoretical calculations 
even in situations where one is not very close to the hadronic threshold 
$s=m_H^2$. As a result, the theoretical uncertainties from perturbation 
theory for the Higgs cross section at the LHC are reduced to a level of 
about 10\%, sufficient for Higgs discovery.

Motivated by all of the above, we will study in the present paper the cross 
section for Higgs production at large transverse momentum $p_T$, typically 
$m_H<p_T <{\mathrm{few}}\times m_H$, and perform the resummation of threshold 
logarithms at NLL accuracy. Ideally, as explained earlier, we would have 
in mind here the process $pp\rightarrow H +\mathrm{jet} + X$, with an 
observed jet at high transverse momentum that roughly balances that of 
the Higgs boson. For simplicity, we will for now only discuss the more 
inclusive reaction $pp\rightarrow H + X$ at large $p_T$, without 
explicit reference to a jet. Since in most cases a high-$p_T$ Higgs 
will indeed be accompanied by a recoiling jet, 
we expect that this process will share many features with 
$H +\mathrm{jet}$ production, in particular regarding the relevance
of perturbative higher-order QCD corrections and resummation that
we wish to study here.

The LO predictions for single-inclusive Higgs cross section at large
 $p_T$, including the full dependence on the mass of the top quark,
 have been known for some time by
 now~\cite{highptLOellis,highptLObaur}.
Several different NLO calculations~\cite{deFlorian,Anastasiou:2004xq,ravindran,glosser}
 exist within the large-top-mass approximation.
Two of these \cite{deFlorian,Anastasiou:2004xq} used numerical integration 
techniques, while the other two derived and provided analytical 
results \cite{ravindran,glosser}.
As in the fully inclusive case, threshold 
logarithms also dominate the cross section when the transverse momentum of 
the Higgs boson is large, even though they are of a somewhat different form. 
In the $p_T$ distribution, when the cross section is integrated over all 
rapidities of the Higgs, they occur in the partonic cross sections as 
$\as^k \ln^m(1-\yTh^2)$, $m\leq 2k$, where $\yTh=(p_T+m_T)/\sqrt{\sh}$ 
with $m_T=\sqrt{\ptsq + m_T^2}$. Also, unlike the fully-inclusive case,
for a Higgs produced at large $p_T$ there needs to be a recoiling parton
already in the Born process, whose color charge plays a role for the structure
of the resummed expression. As is customarily 
done, we will treat the gluon-Higgs interaction in the large $m_t$ limit, 
i.e. by replacing the top loop with an effective $ggH$ coupling. 
Even though this
approximation is not as accurate at large transverse momentum as in case 
of the fully inclusive cross section \cite{Smith:2005yq}, it is 
certainly expected 
to be good for the ratio between higher order calculations and the Born term, 
because the dominant large logarithms are completely independent of the 
structure 
of the $ggH$ coupling. 

We note in passing that kinematically, and conceptually, 
the resummation of the Higgs cross section at large $p_T$ is close to 
that for high-$p_T$ $W$ or $Z$ production in hadronic collisions, considered 
in \cite{Kidonakis:1999ur}. Besides the obvious differences related to
the different final state considered, we also differ from
Ref.~\cite{Kidonakis:1999ur} in our technical treatment of the resummed
formulas. In Ref.~\cite{Kidonakis:1999ur} a NNLO expansion of
the resummed expression is obtained and used, while in the present work
we keep the full NLL-resummed expression. This, as we shall see,
involves a treatment of the whole cross section in Mellin-moment space. 
We also emphasize that the logarithms we are resumming are different from 
those occurring at {\it small} $p_T$ ($p_T\ll m_H$), which have 
received much attention in the 
literature \cite{Hinchliffe:1988ap,ptresum,Bozzi:2005wk,higgsjoint}
since the bulk of the inclusive events is in this regime. Here, the 
resummation has been carried out through NNLL \cite{Bozzi:2005wk}, 
and also a formalism was applied~\cite{higgsjoint} that allows a 
NLL resummation of the logarithms at low $p_T$ jointly with the threshold 
logarithms present in the inclusive ($p_T$-integrated) Higgs 
cross section. Finally, we mention that for a very light Higgs and/or 
at high $p_T$, $p_T\gg m_H$, yet another class of logarithms could become
important, arising through ``fragmentation'' production of the Higgs 
by a final-state gluon. The logarithms in this kinematic regime
have been studied in~\cite{Berger:2001wr} for the Drell-Yan process.
They are not relevant in the threshold situation we are considering
in this work, for which typically $m_H<p_T <{\mathrm{few}}\times m_H$
(and $\hat{y}_T\sim 1$).

The paper is organized as follows: in Section~\ref{sec:cross} we
discuss the structure of the expressions for the Higgs $p_T$ distribution
in fixed-order perturbation theory and discuss the role of the
threshold region. Section~\ref{sec:res} is concerned with the analytical 
results for the threshold-resummed distribution in Mellin-moment space. 
We also describe there the matching of the resummed to the 
fixed-order result, and the prescription for the inverse Mellin transform.
Finally, in Section~\ref{sec:numerics} the phenomenological effects of
threshold resummation on the Higgs $p_T$ distribution at the LHC are 
studied.

\section{Perturbative cross section }
\label{sec:cross}

We consider Higgs production in hadronic collisions, 
\begin{align}
h_1 + h_2 \rightarrow H  + X \, ,
\end{align}
at large transverse momentum $p_T$ of the Higgs boson $H$.
The factorized cross section, differential in $p_T$ and the Higgs rapidity 
$y_H$, can be written as
\begin{align}
\label{eq:1}
\f{ d\sigma}{dp_T^2 dy_H} = \sum_{a,b}\, &
\int_0^1 dx_1 \, f_{a/h_1}\left(x_1,\mu_{F}^2\right) \,
\int_0^1 dx_2 \, f_{b/h_2}\left(x_2,\mu_{F}^2\right) \, 
\f{d\hat{\sigma}_{ab}}{{d{p}_T^2 d{y_H}}} \, ,
\end{align}
where the perturbative partonic cross section is expanded as 
\be
{d \sigh_{ab}  \over  d\ptsq d\yH} = {\sig_0 \over \hat s} \left[
\astpi G_{ab}^{(1)} + \left(\astpi\right)^2 G_{ab}^{(2)} + \dots \right]
\ee
with the partonic center-of-mass energy $\sh = s x_1 x_2$. The Born
cross section, computed within the large-top-mass approximation, is
given by   
\be
\sig_0 = {\pi \over 64} \left( {\as \over 3 \pi v} \right)^2
\ee
with $v$ representing
 the Higgs vacuum expectation value,
 $v^2= 1/(\sqrt 2 G_F)$. In the expressions above,
 $\mu_{F}$ is the factorization scale and the coupling constant
 $\as\equiv \as(\mu_R^2)$ is computed at the renormalization scale
 $\mu_R$. Explicit expressions for the (factorization and renormalization scale
 independent) LO coefficients $G_{ab}^{(1)}$ and the (scale-dependent)
 NLO contributions $G_{ab}^{(2)}$ can be found in~\cite{glosser}. 

The following three partonic channels contribute to this process at the 
lowest order: \mbox{$gg\rar g H$}, $gq
\rar q H$, $q \bar q \rar g H$, the first one being dominant -- as it is
to be expected due to the large gluon-gluon
luminosity at hadron colliders.

In this paper we will for simplicity focus just on the transverse 
momentum distribution of the Higgs boson and integrate over the full range 
of allowed rapidities
\be
\f{ d\sigma}{dp_T^2 } = \int_{y_H^-}^{y_H^+} dy_H
 \f{ d\sigma}{dp_T^2 dy_H} \, ,
\ee
where 
\be
y_H^+=-y_H^-=\f{1}{2} \ln\f{ (1+\sqrt{1-4 s\, \mT^2/(s+m_H^2)^2})}{
  (1-\sqrt{1-4 s\, \mT^2/(s+m_H^2)^2}) }\;,
\ee
with $\mT=\sqrt{m_H^2+\pt^2}$ denoting the transverse mass. 
In our calculation we will express the $\pt$ distribution as a function of
the hadronic threshold variable $\yT$ defined as
\be
\yT= {\pt+\mT \over \sqrt s} \;,
\ee 
i.e. $ { d\sigma}/{dp_T^2 } =  { d\sigma}/{dp_T^2 } (\yT)$.
The limit $\yT \rar 1$ represents the hadronic threshold, i.e., when the
hadronic center-of-mass energy is just enough to produce the Higgs boson 
with a given transverse momentum.

Using the expressions for  $G_{ab}^{(1)}$ in~\cite{glosser} we obtain
the (rapidity integrated) {\it partonic} cross
 sections at the lowest order:
\be
\label{eq:lo}
 { d \sigh_{ab}^{\rm (1)} \over d \ptsq} = \sig_0 \astpi  
\frac{{\cal N}_{ab}(\yTh,r) }{\pt^2\,{\sqrt{1 - \yTh^2}}\,} \, ,
\ee
where the partonic threshold variable $\yTh$ is defined as 
 $\yTh=  \yT/\sqrt{x_1 x_2}$. The square-root factor in the denominator
is a Jacobean from the rapidity integration. 
The coefficients ${\cal N}_{ij}(\yTh,r)$ are 
regular at $\yTh =1$. They also depend on the ``fixed'' quantity 
$r\equiv \pt/m_T$ and are given in Appendix~\ref{app:lo}.

At the next-to-leading order, the integration over rapidity of the
term $G_{ab}^{(2)}$  leads to an expression for the partonic
cross section that can be written as
\be
\label{eq:sv1}
 { d \sigh_{ab}^{\rm (2)} \over d \ptsq} = \astpi  { d \sigh_{ab}^{\rm (1)} 
\over d \ptsq} \left[g_{2,ab}(\pt) \ln^2(1-\yTh^2) + g_{1,ab}(\pt) \ln(1-\yTh^2) + g_{0,ab}(\pt)
\right] + f_{ab}(p_T,\yTh)\, .
\ee
The function $f_{ab}(p_T,\yTh)$ represents terms that vanish in the
limit $\yTh \rar 1$. 

As we discussed in the Introduction, at the $k$th 
order of perturbation theory for the $\sigh_{ij}$, there are 
logarithmically-enhanced contributions of the form
$\as^k \, \ln^m (1-\yTh^2)$, with $m\leq 2k$. In analogy with the inclusive total
Higgs cross section, these logarithmic terms are due to soft-gluon radiation 
and dominate the perturbative expansion when the process is
kinematically close to the partonic threshold. We emphasize that 
$\hat{y}_T$ assumes particularly
large values when the partonic momentum fractions approach
the lower ends of their ranges. Since the parton distributions
rise steeply towards small argument,
this generally increases the relevance of the threshold
regime, and the soft-gluon effects are relevant even for situations
where  the hadronic center-of-mass energy is much larger than the
produced transverse mass of the final state.
For this particular process at the LHC, it has been explicitly checked 
in \cite{Smith:2005yq} that an approximation based on setting 
$f_{ab}(p_T,\yTh)=0$ in Eq.~(\ref{eq:sv1}) gives the 
bulk of the NLO contribution. In the following, we discuss the
resummation of the large logarithmic corrections to all orders in $\as$.

\section{Resummation }
\label{sec:res}

The resummation of the soft-gluon contributions is carried out in
Mellin-$N$ moment space, where the convolutions in Eq.~(\ref{eq:1})
between parton
distributions and subprocess cross sections
factorize into ordinary products. We take Mellin moments in the scaling
variable $y_T^2$ as
\be
\int_0^1 d \yT^2 {(\yT^2)}^{N-1} { d \sig \over d \ptsq} = \sum_{a,b}
f_a (N+1, \mu^2_F) f_b (N+1, \mu^2_F) \sigh_{ab }(N) \; ,
\ee
where the corresponding moments of the partonic cross sections are
\be
\sigh_{ab}(N) = \int_0^1 d \yTh^2 {(\yTh^2)}^{N-1}  
{ d \sigh_{ab} \over d \ptsq} \, ,
\label{partmom}
\ee
and the $f_{a,b} (N+1, \mu^2_F)$ are the usual moments
of the parton distributions in their momentum fractions.
The threshold limit $\hat{y}_T^2\to 1$ corresponds 
to $N\to \infty$, and the leading soft-gluon corrections arise as 
terms $\propto \as^k \ln^{2k}N$. The NLL resummation procedure
discussed in this work deals with the ``towers'' $\as^k \ln^{m}N$ 
for $m=2k,2k-1,2k-2$.

\subsection{Resummation to NLL}
\label{subsec:nll}

In Mellin-moment space, threshold resummation results in
exponentiation of the soft-gluon corrections. In case of the
Higgs cross section at high $p_T$, the resummed cross section 
reads~\cite{threshres,Catani:1996yz}:
\begin{align}
\label{eq:res}
\sigma^{{\rm (res)}}_{ab\to cH} (N-1)=  C_{ab\to cH}\,
\Delta^a_N\, \Delta^b_N\, J_N^c
\Delta^{{\rm (int)} ab\rightarrow cH}_{ N} \,
\sigma^{(1)}_{ab\to cH} (N-1) \;  .
\end{align}
Each of the ``radiative factors'' $\Delta_N^{a,b}$, $J_N^c$, $\Delta^{{\rm (int)} 
ab\rightarrow cH}_{N}$ is an exponential. The factors $\Delta^{a,b}_N$ represent the 
effects of soft-gluon radiation collinear to initial partons $a$ and $b$. 
The function $J^{c}_N$ embodies collinear, soft or hard, emission by the 
non-observed parton $c$ that recoils against the Higgs. 
Large-angle soft-gluon emission is accounted for by the factors
$\Delta^{{\rm (int)} ab\rightarrow cH}_{N}$, which, at variance with the
{\it universal} $\Delta^{a,b}_N$ and $J^{c}_N$ functions, depend on 
the partonic process under consideration.
Finally, the coefficients $C_{ab\to cH}$ contain
$N$-independent hard contributions arising from one-loop
virtual corrections and non-logarithmic soft corrections.
As we mentioned earlier, the structure of the resummed expression
is similar to that for the large-$p_T$ $W$ production cross 
section \cite{Kidonakis:1999ur} or, in the massless limit, 
to that for prompt-photon production in hadronic collisions \cite{photon}.

The expressions for the radiative factors are
\ba
\ln \Delta^{a}_N &=&  \int_0^1 dz \,\f{z^{N-1}-1}{1-z}
\int_{\mu_{F}^2}^{(1-z)^2 Q^2} \f{dq^2}{q^2} A_a(\as(q^2))\, , \nn \\
\ln J^{a}_N &=&  \int_0^1 dz \,\f{z^{N-1}-1}{1-z} \Bigg[ \int_{(1-z)^2
    Q^2}^{(1-z) Q^2} \f{dq^2}{q^2} A_a(\as(q^2)) +\f{1}{2}
  B_a(\as((1-z)Q^2)) \Bigg]\, , \nn \\
\ln\Delta^{(int) ab\rightarrow cH}_{ N} &=& 
 \int_0^1 dz \,\f{z^{N-1}-1}{1-z}\, D_{ ab\to cH}(\as((1-z)^2 Q^2))\, .
\ea
The relevant scale $Q$ for this process is given by  \mbox{$Q^2=p_T^2
(1+r)/r$}. The coefficients ${\mathcal C}= A_a,\, B_a,\,D_{ ab\to cH}$ each 
are a power series in the coupling constant $\as$,
${\mathcal C}= \sum_{i=1}^{\infty}(\as / \pi)^i{\mathcal C}^{(i)}$.
The universal LL and NLL coefficients $A_a^{(1)}$, $A_a^{(2)}$ and  $B_a^{(1)}$ 
are well known \cite{KT,Catani:vd}:
\begin{equation} 
\label{A12coef} 
A_a^{(1)}= C_a
\;,\;\;\;\; A_a^{(2)}=\frac{1}{2} \; C_a K \;,\;\;\;\; B_a^{(1)}=\gamma_a
\end{equation} 
with
\begin{equation} 
\label{kcoef} 
K = C_A \left( \frac{67}{18} - \frac{\pi^2}{6} \right)  
- \frac{5}{9} N_f \;, 
\end{equation}
where $C_g=C_A=N_c=3$, $C_q=C_F=(N_c^2-1)/2N_c=4/3$, $\gamma_q=-3/2 C_F=-2$
 and $\gamma_g=-2\pi \b0$. Here, 
$\b0$ is the first coefficient of the QCD $\beta$-function :
\ba
\b0 &= \frac{1}{12 \pi} \left( 11 C_A - 2 N_f \right) \; .
\label{bcoef}
\ea

The process-dependent coefficient ${D_{ ab \to c H}^{(1)}}$ 
can be obtained either by expanding the resummed formula in Eq.~(\ref{eq:res})
to first order in $\as$ and comparing to the fixed-order NLO result 
in \cite{glosser}, or by explicit computation as outlined in
\cite{Bonciani:2003nt}. We have checked that both approaches result in
\ba
{D_{ ab \to c H}^{(1)}}= (C_a +C_b -C_c) \log\frac{r+1}{r}\, .
\ea
The coefficient is evidently just proportional to a combination of the 
color factors for each hard parton participating in the process.
This simplicity is due to the fact that there is only one color structure
for a process with only three external partons. In the ``massless'' limit
$r \rar 1$ we recover the known expression for the case of prompt-photon 
production \cite{photon}.

The final ingredients for the resummed cross section~(\ref{eq:res}) 
are the lowest order partonic cross sections in Mellin-moment space, 
$\sigma^{{\rm (res)}}_{ab\to cH} (N-1)$, and the coefficients $C_{ ab
\to c H}$. The expressions for the former are presented in 
Appendix~\ref{app:lo}.  Regarding the latter,
at NLL accuracy, we only need to know the first-order term
in the expansion $C_{ ab \to c H} = 1+\sum_{i=1}^{\infty} (\as / \pi)^i
C_{ ab \to cH}^{(i)}$. We derive it by comparing the 
expansion of the resummed expression in Eq.~(\ref{eq:res}) with the 
fixed-order NLO calculation in \cite{glosser}, after going to moment
space. Our results for the one-loop coefficients 
$C_{ ab \to c H}^{(1)}$ are listed in Appendix~\ref{app:coeff}.   

In order to organize the resummation according to the logarithmic accuracy 
of the Sudakov exponents it is customary to expand the latter as 
\ba
\label{lndeltams}
\!\!\! 
\ln \Delta_N^a(\as(\mu_R^2),Q^2/\mu_R^2;Q^2/\mu_{F}^2) 
&\!\!=\!\!& \ln N \;h_a^{(1)}(\lambda) +
h_a^{(2)}(\lambda,Q^2/\mu_R^2;Q^2/\mu_{F}^2) + 
{\cal O}\left(\as(\as \ln N)^k\right) \,, \nn \\ 
\label{lnjfun}
\ln J_N^a(\as(\mu_R^2),Q^2/\mu_R^2) &\!\!=\!\!& \ln N \;f_a^{(1)}(\lambda) +
f_a^{(2)}(\lambda,Q^2/\mu_R^2) + {\cal O}\left(\as(\as \ln N)^k\right)
\,, \nn \\
\label{lnintfun}
\ln\Delta^{(int) ab\rightarrow cH}_{N}(\as(\mu_R^2))
&\!\!=\!\!& \frac{D_{ab \to c H}^{(1)}}{2\pi b_0} \;\ln (1-2\lambda) +
{\cal O}\left(\as(\as \ln N)^k\right) \,,
\ea
with $\lambda=\b0 \as(\mu^2_R) \ln N$. The LL and NLL auxiliary functions
 $h^{(1,2)}$ and $f^{(1,2)}$ are
\ba
\label{h1fun}
h_a^{(1)}(\la)& =&+ \frac{A_a^{(1)}}{2\pi \b0 \la} 
\left[ 2 \la+(1-2 \la)\ln(1-2\la)\right] \;,\\ 
h_a^{(2)}(\la,Q^2/\mu^2_R;Q^2/\mu_{F}^2) 
&=&-\f{A_a^{(2)}}{2\pi^2 \b0^2 } \left[ 2 \la+\ln(1-2\la)\right] - 
\f{A_a^{(1)} \gamma_E}{\pi \b0 } \ln(1-2\la)\nn \\ 
&+& \f{A_a^{(1)} b_1}{2\pi \b0^3} 
\left[2 \la+\ln(1-2\la)+\f{1}{2} \ln^2(1-2\la)\right]\nn \\ 
\label{h2fun}
&+& \f{A_a^{(1)}}{2\pi \b0}\left[2 \la+\ln(1-2\la) \right]  
\ln\f{Q^2}{\mu^2_R}-\f{A_a^{(1)}}{\pi \b0} \,\la \ln\f{Q^2}{\mu^2_{F}} \;,  
\ea
\ba
\label{fll}
f_a^{(1)}(\lambda) =
&-&\frac{A_a^{(1)}}{2\pi b_0 \lambda}\Bigl[(1-2\lambda)
\ln(1-2\lambda)-2(1-\lambda)
\ln(1-\lambda)\Bigr] \; , \\
\label{fnll}
f_a^{(2)}(\lambda,Q^2/\mu^2_R) =
&-&\frac{A_a^{(1)} b_1}{2\pi b_0^3}\Bigl[\ln(1-2\lambda)
-2\ln(1-\lambda)+\f{1}{2}\ln^2(1-2\lambda)-\ln^2(1-\lambda)\Bigr] \nonumber \\
&+&\frac{B_a^{(1)}}{2\pi b_0}\ln(1-\lambda)
-\frac{A_a^{(1)}\GE}{\pi b_0}\Bigl[\ln(1-\lambda)
-\ln(1-2\lambda)\Bigr] \\
&-&\frac{A_a^{(2)}}{2\pi^2 b_0^2}\Bigl[2\ln(1-\lambda)
-\ln(1-2\lambda)\Bigr] 
+ \frac{A_a^{(1)}}{2\pi b_0}\Bigl[2\ln(1-\lambda)
-\ln(1-2\lambda)\Bigr] \ln\frac{Q^2}{\mu^2_R} \; , \nonumber
\ea
where
\ba
b_1=  \frac{1}{24 \pi^2} 
\left( 17 C_A^2 - 5 C_A N_f - 3 C_F N_f \right) \;\;.
\label{b1coef}
\ea

\subsection{Matching and inverse Mellin transform}
\label{subsec:match}

When performing the resummation, one of course wants to make full
use of the available fixed-order cross section, which in our case
is NLO. Therefore, it is appropriate to match the resummed result with
the fixed-order expression. This is achieved by expanding the resummed 
cross section to ${\cal O}(\as^2)$, subtracting the expanded result
from the resummed one, and adding the full NLO cross section:
\begin{align}
\label{hires}
\f{d\sigma^{\rm (match)}(y_T)}{dp_T^2} &= \sum_{a,b}\,
\;\int_{C_{MP}-i\infty}^{C_{MP}+i\infty}
\;\frac{dN}{2\pi i} \;\left( y_T^2 \right)^{-N}
\; f_{a/h_1}(N+1,\mu_F^2) \; f_{b/h_2}(N+1,\mu_F^2)
 \nn \\
&\times \left[ \;
\hat{\sigma}^{\rm (res)}_{ab\to cH} (N)
- \left. \hat{\sigma}^{{\rm (res)}}_{ab\to cH} (N)
\right|_{{\cal O}(\as^2)} \, \right] 
+\f{d\sigma^{\rm (NLO)}(y_T)}{dp_T^2} 
 \;\;,
\end{align}
where $\hat{\sigma}^{{\rm (res)}}_{ab\to cH}$ is the resummed cross
section for the partonic channel $ab\to cH$ as given in Eq.~(\ref{eq:res}).
In this way, NLO is taken into account in full, and the soft-gluon 
contributions beyond NLO are resummed to NLL. Any double-counting
of perturbative orders is avoided. 

Since the resummation is achieved in Mellin-moment 
space, one needs an inverse Mellin transform, 
in order to obtain a resummed cross section in $\yT$ space. This
requires a prescription for dealing with the singularities
at $\lambda=1/2$ and $\lambda=1$ in Eqs.~(\ref{lnintfun})-(\ref{fnll}),
which are a manifestation of the singularity in the perturbative strong 
coupling constant at scale $\Lambda_{\mathrm{QCD}}$. We will use
the ``Minimal Prescription'' developed in Ref.~\cite{Catani:1996yz},
which relies on use of the NLL expanded forms 
Eqs.~(\ref{lndeltams})-(\ref{fnll}), and on choosing
a Mellin contour in complex-$N$ space that lies to the {\it left}
of the poles at $\lambda=1/2$ and $\lambda=1$ in the Mellin integrand:
\begin{align}
\label{hadnmin}
\f{d\sigma^{\rm (res)}(y_T)}{dp_T^2} &=
\;\int_{C_{MP}-i\infty}^{C_{MP}+i\infty}
\;\frac{dN}{2\pi i} \;\left( y_T^2 \right)^{-N}
\sigma^{\rm (res)}(N) \; ,
\end{align}
where $b_0\as(\mu_R^2)\ln C_{MP}<1/2$, but all other poles
in the integrand are as usual to the left of the contour. The
result defined by the minimal prescription has the property that 
its perturbative expansion is an asymptotic series that 
has no factorial divergence and therefore
no ``built-in'' power-like ambiguities.

\section{Higgs transverse momentum distribution at the LHC}
\label{sec:numerics}

Having discussed the resummation formulas, we are now ready to present
results for the high-$\pt$ production of Higgs bosons in the
process $pp \rar H+ X$ at the LHC at 
$\sqrt{s}=14$~TeV, choosing $m_H=125$
GeV as an example. In our analysis we use the latest MRST2004 
set~\cite{mrst2004} of parton distribution functions. Unless 
otherwise stated, we fix the the factorization and renormalization scales to
$\mu_F^2=\mu_R^2=p_T^2+m_H^2$. The considered $\pt$ spectrum starts 
above $\pt=80$ GeV where the effects of {\em small transverse momentum 
logarithms} treated 
in \cite{Hinchliffe:1988ap,ptresum,Bozzi:2005wk,higgsjoint} are less important.

\begin{figure}[htb]
\begin{center}
\epsfig{figure=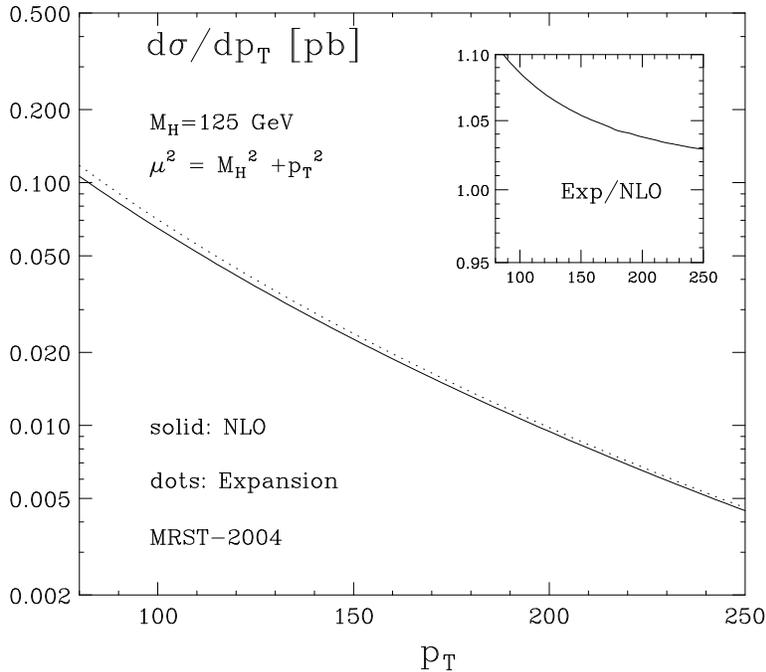,width=0.6\textwidth}
\end{center}
\vspace*{-.6cm}
\caption{ \label{fig1}Comparison between the full NLO result~\cite{glosser} 
and the NLO expansion of the resummed Higgs cross section (corresponding
to the soft-virtual approximation at NLO), at $\sqrt{s}=14$~TeV and
$m_H=125$~GeV. The insert plot shows the 
corresponding ratio.}
\vspace*{-0.2cm}
\end{figure}

First, we confirm that the soft (and virtual) contributions,
corresponding to the terms entering through 
the $g$ functions in Eq.~(\ref{eq:sv1}),  
indeed dominate the cross section. For this we compare the fixed-order 
NLO calculation~\cite{glosser} to the ${\cal O}(\as^2)$ 
expansion of the resummed
expression (the second term in Eq.~(\ref{hires})). Only in the
kinematical region where both contributions are similar can one 
argue that threshold resummation is useful. Figure~\ref{fig1} 
shows the comparison. As can be seen, the soft and virtual terms
faithfully reproduce the full NLO cross section 
to better than 10\% over the whole $p_T$ range considered.  
Towards ``lower'' $p_T$, the agreement deteriorates slightly, 
which is expected since pieces in the cross
section that are not logarithmic in $\yTh^2$
will become more and more important there.  
At very large values of transverse momentum ($p_T>200$
GeV), the process moves kinematically closer to threshold, and
the soft approximation becomes nearly perfect.

\begin{figure}[htb]
\begin{center}
\epsfig{figure=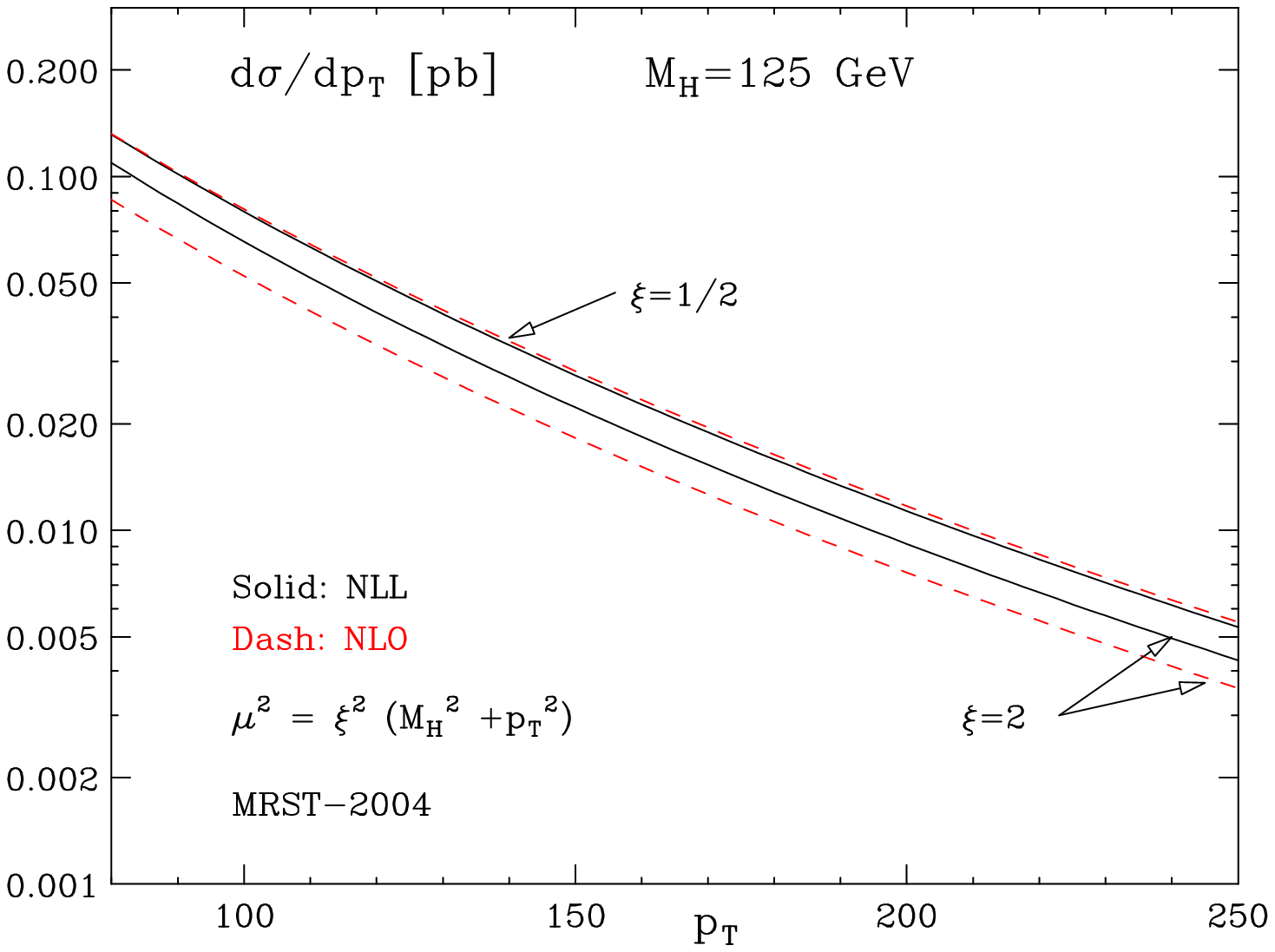,width=0.6\textwidth}
\end{center}
\vspace*{-.6cm}
\caption{ \label{fig2} Scale dependence of the NLO (dashed) and NLL
  (solid) Higgs transverse momentum distributions at the LHC,  
for two different scale choices, $\mu^2=\mu_F^2=\mu_R^2=\xi^2
(p_T^2+m_H^2)$ with $\xi=1/2$ (upper curves) and $\xi=2$ (lower curves). }
\vspace*{-0.2cm}
\end{figure}

One of the virtues of threshold resummation is the reduction of 
the scale dependence of the computed cross sections. For instance, the
scale dependent term $\propto\la \ln(Q^2/\mu^2_{F})$ in Eq.~(\ref{h2fun})
cancels the diagonal part of the DGLAP-evolution of the gluon distribution
at large $N$. To verify this feature for the case of Higgs production we 
show in Fig.~\ref{fig2} the NLO and the NLL resummed (matched) results 
computed for two different values of the scales, $\mu_F^2=\mu_R^2=\xi^2
(p_T^2+m_H^2)$ with $\xi=1/2,2$. A reduction of the scale
dependence by about a factor of two is seen when NLL resummation is 
taken into account. The net effect of the NLL resummation relative
to the NLO cross section, the ``${\rm K}^{{\mathrm{NLL/NLO}}}$-factor''
\ba
\label{k3}
{\rm K}^{{\mathrm{NLL/NLO}}}=\f{d\sigma^{{\mathrm{NLL}}}/dp_T}{
d\sigma^{{\mathrm{NLO}}}/dp_T} \; ,
\ea
is therefore scale dependent. 
While fixed-order and resummed expressions are very similar
for $\xi=1/2$, one finds ${\rm K}^{{\mathrm{NLL/NLO}}}>1$
at larger factorization and renormalization scales.
Overall, we find that threshold resummation does not introduce very
large corrections beyond NLO to the high-$p_T$ Higgs cross section, 
which is somewhat at variance with what was found for the case of 
fully-inclusive Higgs production~\cite{Catani:2003zt}.

\begin{figure}[htb]
\begin{center}
\epsfig{figure=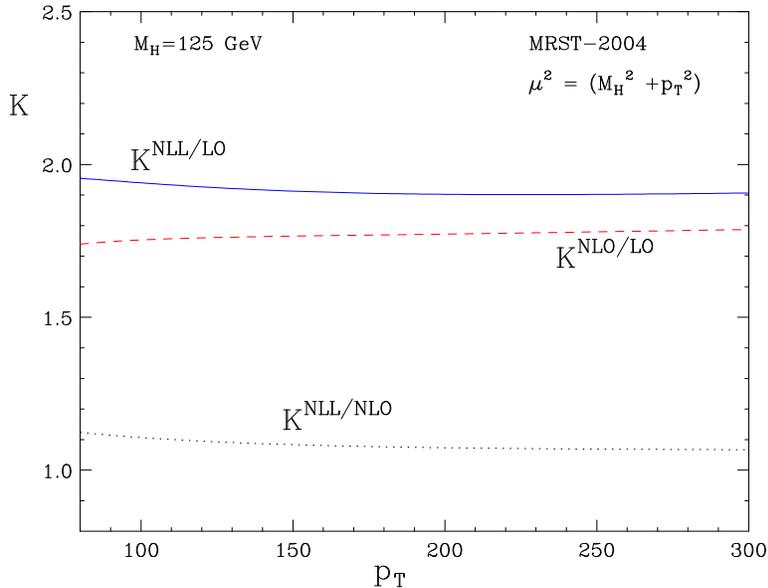,width=0.6\textwidth}
\end{center}
\vspace*{-.6cm}
\caption{ \label{fig3} ``K''-factors for the Higgs transverse momentum
distribution at the LHC, as defined by Eqs.(\ref{k3})-(\ref{k2}).}
\vspace*{-0.2cm}
\end{figure}

We have mentioned before that for the present calculation we 
are using the large-top-mass approximation for the coupling of
two gluons to the Higgs. At large $p_T$, $\pt \gapproxeq m_t$,
this approximation is known to deteriorate~\cite{highptLOellis,highptLObaur}, and a full 
calculation
that includes all effects from the top quark loop will be required%
~\footnote{In order to extend the validity of the results in the soft-virtual
  approximation to larger values of $\pt$, one could 
replace~\cite{Smith:2005yq} the LO
  order cross section calculated in the large $m_t$ limit by the known
  LO cross section for arbitrary $m_t$.}. Fortunately, the 
large logarithms 
we are resumming are insensitive to the structure of the Higgs-gluon 
coupling since they are associated only with emission of soft and 
collinear gluons from the external lines. Therefore, even though
our cross sections shown in Fig.~\ref{fig2} will not be good
predictions anymore at large $p_T$, we can be confident that 
${\rm K}$-factors 
generally will be. In other words, the product between the full 
Born cross section (including all effects from the heavy quark loop) 
as derived in~\cite{highptLOellis,highptLObaur} and our calculated 
${\rm K}$-factors
\ba
\label{k1}
{\rm K}^{{\mathrm{NLO/LO}}}=\f{d\sigma^{{\mathrm{NLO}}}/dp_T}{
d\sigma^{{\mathrm{LO}}}/dp_T}
\ea
and
\ba
\label{k2}
{\rm K}^{{\mathrm{NLL/LO}}}=\f{d\sigma^{{\mathrm{NLL}}}/dp_T}{
d\sigma^{{\mathrm{LO}}}/dp_T}
\ea
should provide a reliable description of the full NLO and NLL 
cross sections. In Fig.~\ref{fig3} we present these ${\rm K}$-factors
along with ${\rm K}^{{\mathrm{NLL/NLO}}}$ for our default 
scale choice $\mu=\sqrt{p_T^2+m_H^2}$. Here
the LO result is obtained using the corresponding MRST LO set of
parton distributions \cite{Martin:2002dr} and the one-loop expression for the 
strong coupling constant. As can be seen from
the dotted line for ${\rm K}^{{\mathrm{NLL/NLO}}}$, resummation 
predicts an increase of about $10\%$ 
of the cross section beyond NLO.
The results presented in Fig.~\ref{fig3} should be taken into 
account in the analysis of future LHC data.


We finally recall that for our predictions we have integrated
over all rapidities of the Higgs. The dependence on rapidity could be 
taken into account in the resummation using the techniques developed 
in~\cite{Sterman:2000pt} for the case of prompt-photon production. 
In that study it was found that the higher-order corrections show 
very little dependence on rapidity unless one considers situations 
with very forward or backward production. We expect the same to be true
for the present case. The ${\rm K}$-factors shown in Fig.~\ref{fig3} 
will therefore also apply to the case where the cross section
is integrated over a finite bin at central rapidities, for example.

\section{ Conclusions and Summary}

The process $pp \rar H (\rar \gamma \gamma) +X$ offers
an enticing possibility of improving the signal-to-background ratio
for Higgs detection at the LHC. In this work we have studied the NLL
resummation of the logarithmic threshold corrections to the partonic
cross sections relevant for this process. We
have presented analytical expressions for the resummed cross section 
in Mellin-moment space. In particular, we have derived the
process-dependent perturbative coefficients necessary for the NLL
resummation. We report a correction of ${\mathcal{O}}(10\%)$ to the
NLO $\pt$ distribution in the range 80 GeV $<\pt<$ 300 GeV for $M_H =
125$ GeV. The resummed result exhibits less dependence on the
factorization and renormalization scales than the NLO 
cross section, implying a reduction of the theoretical 
uncertainties for this process.
\\

\noindent
{\large \bf Acknowledgments} 

\noindent
We are grateful to G.\ Sterman for valuable discussions and for
reading the manuscript, and to M.\ Grazzini for comments on the manuscript.
The work of A.K. was supported by
the Deutsche Forschungsgemeinschaft
in the Sonderforschungsbereich/Transregio SFB/TR-9
``Computational Particle Physics'' and BMBF Grant No. 05HT4VKA/3.
  The work of D.dF. was supported in part
by Fundaci\'on Antorchas, CONICET and UBACyT. 
W.V.\ is grateful to RIKEN, Brookhaven National Laboratory
and the U.S.\ Department of Energy (contract number DE-AC02-98CH10886) for
providing the facilities essential for the completion of his work.

\begin{appendix}

\section{ LO cross sections}
\label{app:lo}

Using the variable $r\equiv \pt/m_T$ , the coefficients of the 
LO cross sections after rapidity integration in 
Eq.~(\ref{eq:lo}) are given in terms of 
${\cal N'}_{ab}(\yTh,r) \equiv {\cal N}_{ab}(\yTh,r)\, \left(1+r\right)^3\,
{\sqrt{{\left( 1 + r \right) }^2 - {\left( 1 - r \right)}^2\,\yTh^2}}
$ as:
\ba
{\cal N'}_{gg}(\yTh,r) &=& 
4 N_c \left({\left( 1 + r \right) }^4 - 2\,{\left( 1 + r \right) }^2\,\yTh^2 + 
      \left( 3 - 2\,r^2 \right) \,\yTh^4 - 2\,{\left( 1 - r \right)
      }^2\,\yTh^6+ {\left( 1 - r \right) }^4\,\yTh^8 \right)\, , \nn \\
{\cal N'}_{gq}(\yTh,r) &=& C_F\,(r+1) \left( 2(1+r)^3 
- 
\left( 1+r\right)\left( 2+r\right)\left( 2-r\right)
 \,\yTh^2 +
      3\,\left( 1 - r \right) \,\yTh^4  \right. \nn \\
&-& \left. {\left( 1 - r \right) }^3\,\yTh^6 \right)\, , \nn \\
{\cal N'}_{gq}(\yTh,r) &=&{\cal N'}_{qg}(\yTh,r)\, , \nn \\ 
{\cal N'}_{q \bar q} (\yTh,r)&=& 4 C_F^2 r^2 \yTh^2
    \left( {\left( 1 + r \right) }^2 - 2 \yTh^2 + {\left( 1 - r
 \right) }^2 \yTh^4 \right) \, .
\ea

The explicit expressions for the Mellin moments of the LO
partonic cross sections are: 
\ba
\sigh^{(1)}_{gg \to g H}(N) &=&\frac{2 \as N_c \sig_0}{\sqrt{\pi } 
\pt^2 {\left( 1 + r \right) }^4} 
\left[ \frac{{\left( 1 + r \right) }^4 {\cal F}_N(0,z)
  \Gamma(N)}{\Gamma(\frac{1}{2} + N)}
- \frac{2 {\left( 1 + r \right) }^2 {\cal F}_N(1,z) \Gamma(1
    +N)}{\Gamma(\frac{3}{2} + N)}
 \right. \nn \\ 
&+& \left. \frac{\left( 3 - 2 r^2 \right)  {\cal F}_N(2,z) 
\Gamma(2 +N)}{\Gamma(\frac{5}{2} + N)} 
- \frac{2 {\left( 1 - r \right) }^2 {\cal F}_N(3,z) 
\Gamma(3 + N)}{\Gamma(\frac{7}{2} + N)}  \right. \nn \\ 
&+& \left. \frac{{\left( 1 - r \right)}^4 {\cal F}_N(4,z) \Gamma(4
  +N)}{\Gamma(\frac{9}{2} + N)} \right] \, ,
\nn \\
\sigh^{(1)}_{gq \to q H}(N) &=& \frac{\as C_F \sig_0 }{2 
\sqrt{\pi }\pt^2 {\left( 1 + r \right) }^3} 
\left[ \frac{2 {\left( 1 + r \right) }^3 {\cal
      F}_N(0,z)\Gamma(N)}{\Gamma(\frac{1}{2} + N)} \right. \nn \\
&-& \left. \frac{\left( 1+r\right)\left( 2+r\right)
\left( 2-r\right){\cal F}_N(1,z) \Gamma(1
  +N)}{\Gamma(\frac{3}{2} + N)}  \right. \nn \\ 
&+& \left. \frac{3 \left( 1 - r \right)  {\cal F}_N(2,z)
\Gamma(2 + N)}{\Gamma(\frac{5}{2} + N)} 
- \frac{{\left( 1 - r \right)}^3 {\cal F}_N(3,z) 
\Gamma(3+N)}{\Gamma(\frac{7}{2} + N)} \right] \, ,
\nn \\
\sigh^{(1)}_{q \bar q \to g H}(N) &=&\frac{2 \as C_F^2 r^2 
\sig_0}{\sqrt{\pi}\pt^2{\left( 1+ r \right)}^4}  
\left[ \frac{{\left( 1 + r \right) }^2 {\cal F}_N(1,z) 
\Gamma(1 + N)}{\Gamma(\frac{3}{2} + N)} 
- \frac{2 {\cal F}_N(2,z) \Gamma(2 + N)}{\Gamma(\frac{5}{2} + N)}  
\right. \nn \\ 
&+& \left.
 \frac{{\left( 1 - r \right) }^2 {\cal F}_N(3,z) 
\Gamma(3 + N)}{\Gamma(\frac{7}{2} + N)} \right]\, ,  \nn \\ 
\sigh^{(1)}_{qg \to qH}(N) &=&\sigh^{\rm (1)}_{gq \to qH}(N) \, ,
\ea
where ${\cal F}_N(n,z) \equiv \!\!\!\phantom{x}_2 
F_1({1 / 2},N+n,N+ (n+1)/2; z)$ and $z \equiv (r-1)^2 / (r+1)^2$. For
large $\pt$, the variable $r$ is close to $1$, and for numerical 
purposes it is therefore sufficient
to expand the Hypergeometric function $\!\!\!\phantom{x}_2 
F_1$ to second order in $z$:
\be
\!\!\!\phantom{x}_2F_1 (a,b,c;z) = 1+ { a b \over c} z + { a (a+1)
  b (b+1) \over 2 c (c+1)} z^2 + {\cal O}(z^3) \; .
\ee

\section{One loop coefficients }
\label{app:coeff}
The one-loop coefficients $C_{ab \to c H}^{(1)}$ for the three different 
subprocesses read
\ba
C_{gg \to g H}^{(1)}&=& \frac{11}{2} + \frac{16\,C_A}{9} - 
\frac{7\,N_f}{36} + \frac{5\,C_A\,{\pi }^2}{12} + 
\pi \b0 \GE + \f{3 C_A }{2}\GE^2\nn \\
&+&\left(C_A-N_f\right) \f{1-2 r+10r^2}{12\left(1+6 r^2+2 r^4\right)} 
+ 2\,C_A\,{\rm Li}_2(1 - r) + 
  C_A\,{\rm Li}_2\left(\frac{2\,r}{1 + r}\right) \nn \\
&+& 
  2\,C_A\,\ln (1 - r)\,\ln r - \frac{C_A}{2}\ln^2 r - 
C_A\,\ln r\,\ln (1 + r) + 
  \frac{C_A}{2}\ln^2 (1+r)\nn \\
&-& C_A \GE \ln\f{1+r}{r}  
+ 2 \left(\pi b_0 -C_A \GE\right)\,\ln \frac{Q^2}{\mu_F^2 }
 -  3 \,\pi b_0\,\ln \f{Q^2}{\mu_R^2 }\, ,
\ea
\ba
C_{gq \to q H}^{(1)} &=& \frac{11}{2} - \frac{9\,C_F}{4} + 
\frac{38\,C_A}{9} - \frac{5\,N_f}{9} - \frac{C_F\,{\pi }^2}{4} + 
  \frac{2\,C_A\,{\pi }^2}{3} + \frac{\left(C_A-C_F\right)\,r}{2\,\left( 1 +
    2\,r\,\left( 1 + r \right)  \right) } \nn \\
&+& \GE^2 \left(C_A+\f{C_F}{2}\right) +\f{3}{4} \GE C_F -C_A \GE 
\ln \f{1+r}{r} +  \left( C_F + C_A\right)\ln (1 - r)\,\ln r
\nn \\
&-& 
 \f{C_A}{2} \ln^2 r 
- C_F\,\ln r\,\ln (1 + r) + \frac{C_F}{2}\,{\ln^2 (1 + r)} 
+\left(C_F+C_A\right) \,{\rm Li}_2(1 - r) \nn \\ 
&+& C_F\,{\rm Li}_2\left(\frac{2\,r}{1 + r}\right) + \left( \pi b_0 + 
\f{3}{4}C_F -C_F \GE - C_A \GE \right)\ln \frac{Q^2}{\mu_F^2 }
-3 \,\pi b_0\,\ln \frac{Q^2}{\mu_R^2 }\, ,
\ea

\ba
C_{q \bar q \to g H}^{(1)} &=&\frac{11}{2} - \frac{9\,C_F}{2} +
\frac{79\,C_A}{12} - \frac{5\,N_f }{6} + \frac{4\,C_F\,{\pi }^2}{3} - 
  \frac{11\,C_A\,{\pi }^2}{12} + \frac{C_A-C_F}{2\,r}+\GE\pi\b0 \nn \\
&+& 
  2\,C_F\,\ln (1 - r)\,\ln r +\f{C_A}{2}\ln^2\frac{1+r}{r}
- C_F \ln^2 r + \left( \f{3}{2}C_F-2\pi\b0  \right) \ln
\frac{1+r}{r} \nn \\
&+&2\,C_F \,{\rm Li}_2(1 - r) + 
  C_A\,{\rm Li}_2\left(\frac{2\,r}{1 + r}\right)-3\pi\b0\,
\ln \frac{Q^2}{\mu_R^2 } \nn \\
&+&\left(C_A-2 C_F \right)\GE \ln \f{1+r}{r}  +\GE^2 \left(2C_F-
\f{C_A}{2}\right)
 + C_F \left(\f{3}{2}-2 \GE \right)\ln \frac{Q^2}{\mu_F^2 }\, ,
\ea
where $\b0$ is given in Eq.~(\ref{bcoef}).
\end{appendix}



\end{document}